\documentclass[aps,showpacs,pra]{revtex4}
\usepackage{amsfonts}
\usepackage{amsmath}
\usepackage{amssymb}
\usepackage{graphicx}

\begin{document}

\preprint{}
\title{Collective excitations of low density fermion-boson quantum-liquid
mixtures}
\author{D. H. Santamore}
\affiliation{ITAMP, Harvard-Smithsonian Center for Astrophysics,
Cambridge, Massachusetts 02138} \affiliation{Department of Physics,
Harvard University, Cambridge, Massachusetts 02138}
\author{Eddy Timmermans}
\affiliation{T-4, Theory division, Los Alamos National Laboratory,
Los Alamos, New Mexico 87545} 

\begin{abstract}
We investigate the collective excitations of a low temperature
dilute gas mixture that consists of a Bose-Einstein condensate and a
Fermi-gas that is a normal (i.e. non-superfluid) Fermi-liquid. We
find that the BEC-mediated fermion-fermion interactions, as a
consequence of retardation, can become repulsive and support a
zero-sound mode that is essentially undamped. In addition, we find a
damped zero-sound mode that can be described as a BEC-sound mode
modified by fermion mediated boson-boson interactions, and we derive
its decay-rate caused by Landau damping. We study the mode structure
of these excitations and find avoided crossing behavior as well as a
termination point. The collective mode dynamics also reveals that
phase separation sets in when the fermion-mediated boson-boson
interaction destroys the stability of the homogeneous BEC. We
estimate the time and length scales of the onset of the phase
separation, and we discuss the feasibility of experimentally probing
these consequences of mediated interactions.
\end{abstract}

\date{\today}
\maketitle

\section{Introduction}

Current atom trap experiments create quantum degenerate Fermi-gas
systems \cite{GrimJinKetSalThom,sympat} and Fermi-Bose gas mixtures
in sympathetic cooling experiments \cite{sympat}. The sympathetic
cooling of fermion atoms utilizes bosonic atoms at lower
temperature, often condensed into a Bose-Einstein condensate (BEC).
At the time of writing, the fermions do not quite enter the
Fermi-liquid regime before the BEC is separated out and the fermions
are further cooled evaporatively \cite{FLregime}; however,
unambiguous Fermi-liquid/BEC mixtures \textit{are} attainable
\cite{Eddy}. Fermi-liquid behavior sets in when the energy can be
expanded up to second order in the (quasi-) particle occupation
numbers, the heat capacity varies linearly with temperature, and the
Pauli exclusion principle sufficiently inhibits incoherent fermion
scattering to validate the assumption of collisionless dynamics. A
dilute two-component Fermi-gas enters this regime when the
temperature drops below $10 \%$ of the Fermi temperature \cite{PN}.
Strong interaction effects further lower the Fermi-liquid range:
experiments reported an upper bound of $2 \%$ of the
Fermi-temperature in the condensed $^{3}$He liquid.

With the creation of cold atom Fermi--Bose quantum liquid mixtures,
atom traps can access novel regimes of low temperature physics. At
present, table-top explorations of Fermi--Bose quantum-liquid
mixtures are confined to the condensed $^{3}$He--$^{4}$He mixtures.
These experiments revealed an intricate phase diagram, even if the
$^{3}$He component did not reach superfluidity. The critical
temperature for $^{4}$He superfluidity drops in the $^{3}$He `rich'
regime, and depending on the temperature and fraction of the
$^{3}$He isotope, the mixture spontaneously breaks translational
symmetry by forming '$^{3}$He only' regions. The microscopic
understanding of this phase separation is, however, complicated due
to the strong interaction which masks or, at least, competes with
the quantum statistical effects \cite{KS93}.

General arguments based on Fermi-liquid theory indicated the
importance of mediated interactions \cite{BBP66}; they are the
modification of the inter-particle interactions in one liquid due to
the presence of the second liquid and directly causes the phase
separation by destroying the stability of the homogeneous state. In
the static limit, applicable to the description of the helium
mixtures in which the sound velocity $c$ of the boson superfluid
greatly exceeds the Fermi velocity $v_{\mathrm{F}}$, the
boson-mediated fermion-fermion interactions are attractive and can
Cooper-pair the fermions \cite{Leeuwen, Pinessup, Lukin}. From the
experimental measurements in helium mixtures, it appears that the
strong interactions greatly weaken the mediated interactions
\cite{BBP66}.

In contrast, the phase separation of cold atom Fermi-mixtures
\cite{Viverit} would be amenable to a transparent first-principles
description \cite{Stoof} and the boson-mediated fermion-fermion
interactions can significantly increase the critical temperature for
Cooper-pairing \cite{Heiselberg}. In addition, the relative
densities of the cold atom mixtures can reverse the
Fermi-velocity $v_{\mathrm{F}}$ to Bose sound velocity $c$ ratio to yield $%
c/v_{\mathrm{F}}\leq1$, which is not attainable in the helium
mixtures. Under those conditions, satisfied in most sympathetic
cooling experiments, the static description breaks down and we show
that retardation radically alters the nature of the boson-mediated
fermion-fermion interactions.

There are other aspects that motivate cold atom Fermi-liquid BEC
studies. For example, the range of the BEC-mediated fermion-fermion
interactions, which is the BEC-healing length, can exceed the
inter-particle distances. Cold atom experiments with fermion-BEC
mixtures can then go beyond the usual cold atom paradigm of
short-range, contact-like inter-particle potentials while examining
similarities with other Fermi-liquids. In particular, the form of
the static BEC-mediated interaction suggests an intriguing
connection between the anticipated cold atom Cooper-paired
Fermi--Bose mixtures and the BEC-exciton superfluid that has been
pursued in condensed matter for more than two decades
\cite{Nozieres}. The BEC-mediated interaction potential is of the
Yukawa form \cite{Peshkin}, which resembles the electron-hole
Coulomb attraction of an exciton at distances smaller than the
Yukawa range.

The realization of s-wave Cooper-pairing in fermion-BEC mixtures,
will, however, require much experimental skill. This pairing scheme
requires the simultaneous trapping of fermion atoms in two distinct
hyperfine states at nearly equal densities, a considerable challenge
in the presence of a BEC: A dependence of the boson-fermion
interactions on the fermion spin introduces a difference in the
local spin densities even if the total number of fermion atoms is
evenly divided over the two spin states and both spins experience
identical trapping potentials.

In this paper, we focus on a dilute mixture of BEC and one-component
Fermi gas that remains normal. We present a theoretical study of the
small amplitude oscillations around a spatially uniform ground
state. First, we show that the resonant dynamics of the mediated
interactions introduces novel features. If $c<v_{\mathrm{F}}$, the
mediated interaction supports a long-lived essentially undamped zero
sound mode as in a two-component Fermi-liquid case, which is
incompatible with the effective attraction predicted by the static
treatment. On the other hand, if $v_{\mathrm{F} }<c$, the mixture
supports an additional mode that is the remnant of the BEC-sound
mode modified by a cloud of fermion excitations that introduce
fermion mediated boson-boson interactions. This second mode is
damped. Both oscillations have been discussed by Yip \cite{Yip} from
a calculation of many-body response functions. Several calculations
of zero-sound modes in trapped fermions have been also reported in
\cite{Capuzzi}. We analyze the homogeneous BEC--fermion mixture
using the semi-classical description, and obtain a single
self-consistent Fermi-liquid-like collective mode equation. The
semiclassical description in terms of the spatially oscillating
deformation of the local Fermi-surface elucidates the zero-sound
nature of the fermion oscillations in both modes, revealing that the
dominant decay mechanism of the BEC-like sound mode at long
wavelengths is Landau damping.

The dispersion relations of the two collective modes, the undamped
and damped modes, avoid crossing each other in the wavenumber
interval where the undamped mode gradually changes its character
from a pure Fermi-liquid zero sound mode to a Bogoliubov excitation
of the BEC. The damped mode terminates at a certain wavelength. The
damping rate at the termination point remains significantly smaller
than the excitation energy so that the mode represents a good
quasi-particle for all wavenumbers and the termination point should
be observable.

The phase separation is manifested by the collective mode dynamics.
The instability corresponds to collective mode eigenvalues of purely
imaginary value. Our analysis reveals the mechanism of the
instability, and gives the length and time scales on which the phase
separation sets in.

First, we discuss the mediated interaction of the fermion--BEC in a
static limit in Sec. II. Then, the dynamical effects are obtained in
Sec.\ III, followed by the mode analysis in Sec.\ IV. Finally,
experimental aspects are examined in Sec.\ V. This is an extension
of the work previously presented in Ref.\ \cite{us}.

\section{Mediated Interactions in the Static limit $\left( \protect\omega
\rightarrow0\right) $\label{Sec_Static}}

In most mixtures, the mediated interaction shifts the frequency of
the existing mode. In the single-component fermion BEC mixture, the
BEC-mediated interactions actually support collective Fermi-liquid
like oscillations. Interactions amongst identical Fermions are
prevented by Pauli-exclusion, so that any fermion-fermion
interactions in the single component fermion BEC mixture are
necessarily BEC mediated.

In this section we illustrate BEC-mediated interactions by
considering the simplest system to exhibit this phenomenon: two
indistinguishable, stationary fermions of mass $m_{\mathrm{F}}$ are
embedded in a dilute BEC of the particle density
$\rho_{\mathrm{B}}^{0}$ and particle mass $m_{\mathrm{B} }$ in
equilibrium located at position $\mathbf{x}_{1}$, and
$\mathbf{x}_{2}$ . The mediated interaction obtained in a static
system highlights the dynamical effects that we discuss in Sec.\
\ref{Sec_collective dynamics}. Without the impurity atoms, the BEC
would be homogeneous. The bosons interact by short range
interactions, which we describe by a pseudo-potential $\lambda
_{\mathrm{BB}}\delta(\mathbf{r}-\mathbf{r} ^{\prime})$ with
interaction strength, $\lambda_{\mathrm{BB}
}=4\pi\hbar^{2}a_{\mathrm{BB}}/m_{\mathrm{B}}$, that is proportional
to the boson-boson scattering length $a_{\mathrm{BB}}$. Similarly,
we assume that the interaction between the fermion and BEC-boson
atoms is well described by
$\sum_{j}\lambda_{\mathrm{BF}}\delta(\mathbf{x}_{j}-\mathbf{r})$,
with an interaction strength proportional to the impurity-boson
scattering length $ a_{\mathrm{BF}}$ and inversely proportional to
the effective mass of the fermion-boson system $m_{\mathrm{BF}}$,
where $m_{\mathrm{BF}}^{-1}=m_{
\mathrm{F}}^{-1}+m_{\mathrm{B}}^{-1}$, $\lambda _{\mathrm{BF}
}=2\pi\hbar^{2}a_{\mathrm{BF}}/m_{\mathrm{BF}}$. The
pseudo-potential description presupposes that the scattering lengths
are shorter than the average inter-particle distances,
$a_{\mathrm{BB}},a_{\mathrm{BF}}\ll\left( \rho_{B}^{0}\right)
^{-1/3}$.

This system lends itself to a particularly transparent description
of the mediated interaction. The BEC-mediated interaction is the
modification of the mean-field energy experienced by the fermion at
$\mathbf{x}_{1}$ due to the other fermion at position
$\mathbf{x}_{2}$. The energy shift is caused by the change in
BEC-density $\rho_{\mathrm{B}}(\mathbf{x}_{1})$ due to the
interaction of the other atom with the BEC. Within the mean-field
description, the equilibrium BEC field in the presence of the
$\mathbf{x}_{2} $ atom follows from the time-independent
Gross-Pitaevskii equation
\begin{equation}
\mu_{\mathrm{B}}\phi(\mathbf{r})=\left[
\frac{-\hbar^{2}\nabla_{\mathbf{r}
}^{2}}{2m_{B}}+\lambda_{BB}|\phi(\mathbf{r})|^{2}\right]
\phi(\mathbf{r}
)+\lambda_{BF}\delta(\mathbf{r}-\mathbf{x}_{2})\phi(\mathbf{r})\;,
\label{tigp}
\end{equation}
where $\phi({\mathbf{r}})$ represents the condensate field
$\phi=\langle \hat{\psi}\rangle$ with
$\rho_{B}(\mathbf{r})=|\phi(\mathbf{r})|^{2}$ and $
\mu_{\mathrm{B}}=\lambda_{\mathrm{BB}}\rho_{\mathrm{B}}^{0}$ is the
chemical potential of the BEC ensuring that the density tends to
$\rho_{\mathrm{B} }^{0}$ at large distances from $\mathbf{x}_{2}$.
Since the BEC is in equilibrium, the field $\phi$ can be chosen to
be real-valued.

Linearizing the condensate wavefunction by setting $\phi \approx
\sqrt{\rho _{\mathrm{B}}^{0}}+\delta \phi $, Eq.(\ref{tigp}) becomes
\begin{equation}
\left[ \nabla _{\mathbf{r}}^{2}-\xi ^{-2}\right] \delta \phi (\mathbf{r}%
)=-\Xi \delta (\mathbf{r}-\mathbf{x}_{2}),  \label{Yk}
\end{equation}%
where $\xi $ is the healing length of the BEC (also referred to as
the coherence length), $\xi =1/\sqrt{16\pi \rho
_{\mathrm{B}}^{0}a_{\mathrm{BB}}}
$ and $\Xi $ is the strength of the source term, $\Xi =-4\pi (1+m_{\mathrm{B}%
}/m_{F})a_{\mathrm{FB}}\sqrt{\rho _{\mathrm{B}}^{0}}$. The solution to Eq.\ (%
\ref{Yk}) is proportional to the Green function of the modified
Helmholtz equation,
\begin{equation}
\delta \phi (\mathbf{r})=-\sqrt{\rho _{\mathrm{B}}^{0}}\left(
1+\frac{m_{ \mathrm{B}}}{m_{\mathrm{F}}}\right)
\frac{a_{\mathrm{FB}}}{|\mathbf{x}_{2}-
\mathbf{r}|}e^{-\frac{|\mathbf{x}_{2}-\mathbf{r}|}{\xi }}.
\end{equation}
The condition, $|\delta \phi \left( \mathbf{r}\right) |/\sqrt{\rho
_{\mathrm{ B}}^{0}}\equiv \zeta \ll 1$ suggests that the Yukawa
profile is accurate if $ \left\vert
\mathbf{x}_{1}-\mathbf{x}_{2}\right\vert >r_{s}$, where $r_{s}\equiv
a_{\mathrm{BB}}(1+m_{\mathrm{B}}/m_{\mathrm{F}})/\zeta$, and we have
assumed that $\xi >r_{s}$. The BEC-mediated fermion--fermion
interaction potential in the static limit is
\begin{align}
V_{\mathrm{F}}^{\mathrm{med}}\left( \left\vert
\mathbf{x}_{1}-\mathbf{x} _{2}\right\vert \right) & =2\lambda
_{\mathrm{FB}}\sqrt{\rho _{\mathrm{B}
}^{0}}\delta \phi (\mathbf{x}_{1})  \notag \\
& =-2\left( 1+\frac{m_{\mathrm{B}}}{m_{\mathrm{F}}}\right) \lambda
_{\mathrm{FB}}\rho _{\mathrm{B}}^{0} a_{\mathrm{FB}}\frac{\exp
\left( \frac{-\left\vert \mathbf{x}_{1}-\mathbf{x}_{2}\right\vert
}{\xi }\right) }{\left\vert \mathbf{
x}_{1}-\mathbf{x}_{2}\right\vert }.  \label{eq:med}
\end{align}
The result is an attractive Yukawa potential, familiar from the
description of effective interactions mediated by massive scalar
fields in relativistic field theory \cite{Peshkin}. For BEC's this
result was obtained as well using perturbation theory \cite{Viverit,
Stoof, PS}.

From Eq.\ (\ref{eq:med}) we find that the average value of the
static mediated interaction for a fermion particle in a homogeneous
BEC-fermion mixture is only a fraction of the mean-field
fermion--boson interaction energy experienced by a fermion, $\lambda
_{\mathrm{FB}}\rho _{\mathrm{B} }^{0}$, with the fraction being of
the order of $2\left( 1+m_{\mathrm{B}}/m_{ \mathrm{F}}\right) \left(
\rho _{\mathrm{F}}^{0}\right) ^{1/3}a_{\mathrm{FB}}
\exp\left(-1/[\xi \rho_{F}^{0})^{1/3}\right)$. However, in the long
wavelength limit, the Fourier transformed potential,
$V_{\mathrm{FF}}^{\mathrm{med}}(k)$, is
\begin{equation}
V_{\mathrm{FF}}^{\mathrm{med}}(k)=-\frac{\lambda
_{\mathrm{FB}}^{2}}{\lambda _{\mathrm{BB}}}\frac{1}{1+\left( k\xi
\right) ^{2}},  \label{eq:vk}
\end{equation}
and comparable to or possibly larger than that of the effective
contact potential, $\lambda _{\mathrm{FB}}$, $\lim_{k\rightarrow
0}V_{\mathrm{FF}}^{ \mathrm{med}}(k)=-\lambda
_{\mathrm{FB}}^{2}/\lambda _{\mathrm{BB}}$. We note that Eq.\
(\ref{eq:vk}) agrees with the long wavelength limit derived in Ref.\
\cite{BBP67} using the Landau Fermi-liquid approach (which is also
valid in describing the strongly interacting systems),
\begin{equation}
V_{\mathrm{FF}}^{\mathrm{med}}(k=0)=\lim_{p,p^{\prime }\rightarrow
0}\frac{
\partial E_{p}}{\partial \rho _{\mathrm{B}}^{0}}\frac{\partial \rho _{
\mathrm{B}}^{0}}{\partial \mu _{B}}\frac{\partial E_{p^{\prime
}}}{\partial \rho _{\mathrm{B}}^{0}}\;,  \label{lfl}
\end{equation}
where $E_{p}$ is the fermion quasi-particle dispersion, and the
derivatives are taken while keeping the occupation numbers constant.
For the weak-interaction, $E_{p}\approx \lambda _{\mathrm{FB}}\rho
_{\mathrm{B} }^{0}+p^{2}/2m_{\mathrm{F}}$, and we recover the
long-wavelength limit of Eq.\ (\ref{eq:vk}).

Similarly, we can determine the fermion-mediated interactions in the
static limit by calculating the density variation caused by a single
impurity atom located at $\mathbf{x}$, interacting with the fermions
via a short-range interactions. The resulting density variation
oscillates on the length scale of the average fermion-fermion
distance, yielding a density profile $\rho _{
\mathrm{F}}(\mathbf{r})$ with features known as
Friedel-oscillations. At large distances, the density variation
$\delta \rho _{\mathrm{F}}(\mathbf{r}) $ of a three-dimensional
weakly interacting fermion system oscillates as
\begin{equation}
\delta \rho _{\mathrm{F}}(\mathbf{r})\propto \frac{\cos
(k_{\mathrm{F}}| \mathbf{x}-\mathbf{r}|+\kappa
)}{|\mathbf{x}-\mathbf{r}|^{3}} \label{Friedel}
\end{equation}%
where $k_{\mathrm{F}}$ is the Fermi momentum, $k_{\mathrm{F}}=(6\pi
^{2}\rho _{\mathrm{F}}^{0})^{1/3}$, and $\kappa $ is the phase
shift. The resulting mediated interaction is known as the
RKKY-interaction.

\section{Collective mode dynamics\label{Sec_collective dynamics}}

While the static limit of the mediated interactions presents a
picture of appealing simplicity, most sympathetic cooling
experiments operate outside the boundary of its validity (which
amounts to $v_{\mathrm{F}}\ll c$). In this section, we investigate
the effects of the dynamics of the BEC-response in collective modes.
The BEC density fluctuation propagates at a finite velocity and the
consequent retardation of its response to a fermion density
fluctuation can alter the physics of the collective oscillations
radically as we show below.

We assume that the boson-boson interactions are repulsive,
$a_{\mathrm{BB}}>0 $, and the temperature is sufficiently low so
that almost all boson particles are condensed into the BEC, thus the
fermion dynamics can be described as collisionless. These
assumptions will lead to a description of the dynamics that is
effectively temperature independent even though it is tacitly
understood that the temperature exceeds the critical temperature for
fermion pairing.

\subsection{Fermion dynamics\label{SubSec_fermion}}

The inhibition of incoherent fermion-fermion scattering by
Pauli-exclusion principle does not eliminate the effects of
fermion-fermion interactions. As shown by Landau \cite{Landauzero},
collective modes can be driven by the mean-field variations that
accompany the propagating density and current variations.
Zero-sound, which corresponds to a periodic deformation of the local
Fermi-surface, is the quintessential example of such motion.

Let us define the single particle fermion distribution function as
$n( \mathbf{r},\mathbf{p};t)$, where $\mathbf{r}$ is the location
and $\mathbf{p} $ is the momentum. The time evolution of this
function is governed by the collisionless transport equation
\begin{equation}
\frac{\partial n\left( \mathbf{r},\mathbf{p},t\right) }{\partial
t}+\mathbf{v }\cdot\frac{\partial n\left(
\mathbf{r},\mathbf{p},t\right) }{\partial
\mathbf{r}}+\mathbf{F}\left( \mathbf{r},\mathbf{p},t\right)
\cdot\frac{
\partial n\left( \mathbf{r},\mathbf{p},t\right) }{\partial \mathbf{p}}=0\;,
\label{eq:dn/dt}
\end{equation}
where we have approximated the velocity of the dilute fermion gas as
$ \mathbf{v}\simeq\mathbf{p}/m_{\mathrm{F}}$, and $\mathbf{F}$ is
the gradient of the BEC-induced mean-field interaction energy,
$\mathbf{F}=-\overline { \nabla}\left[
\lambda_{\mathrm{FB}}\rho_{\mathrm{B}}\left( \mathbf{r} ,t\right)
\right] $, and, in Cartesian coordinates, $\overline {\nabla} =
\left( \frac{\partial}{\partial x}, \frac{\partial}{\partial y},
\frac{\partial}{\partial z}\right) $. Then, we linearize $n$
assuming that $n(\mathbf{r}, \mathbf{p};t)$ fluctuates around the
equilibrium distribution $n^{0}\left( \mathbf{p}\right) $ and the
fluctuation, which is confined in momentum space to a thin shell in
the vicinity of the equilibrium Fermi-surface, propagates
in real space as a plane wave with wavevector $\mathbf{k}$. We write $%
n\left( \mathbf{r},\mathbf{p},t\right) $ as
\begin{equation}
n\left( \mathbf{r},\mathbf{p},t\right) =n^{0}\left( \mathbf{p}\right) +u_{%
\mathbf{p}}p_{\mathrm{F}}\delta\left(
|\mathbf{p}|-p_{\mathrm{F}}\right) e^{i\left(
\mathbf{k}\cdot\mathbf{r}-\omega t\right) },   \label{an}
\end{equation}
where $p_{\mathrm{F}}$ is the equilibrium Fermi momentum and
$u_{\mathbf{p}}$ denotes the amplitude of deformation. Similarly,
assuming that the BEC
fluctuation also propagates as a plane wave, $\rho_{B}(\mathbf{r},t)=\rho_{%
\mathrm{B}}^{0}+\delta\rho_{B}\left( \mathbf{r},t\right) $ with $\delta\rho_{%
\mathrm{B}}\left( \mathbf{r},t\right) =\eta_{\mathbf{k}}\rho_{\mathrm{B}%
}^{0}\exp\left[ i\left( \mathbf{k}\cdot\mathbf{r}-\omega t\right)
\right] $, we obtain from Eq.\ (\ref{eq:dn/dt}) with Eq.(\ref{an})
that
\begin{equation}
-i\omega u_{\mathbf{p}}+i\frac{\mathbf{k}\cdot\mathbf{p}}{m_{\mathrm{F}}}u_{%
\mathbf{p}}=-i\frac{\mathbf{k}\cdot\mathbf{p}}{p_{\mathrm{F}}^{2}}\lambda_{%
\mathrm{FB}}\rho_{\mathrm{B}}^{0}\eta_{\mathbf{k}},
\label{eq:fermiondynamics}
\end{equation}
In the plane wave propagation of the BEC fluctuation, the complex
phase of
the BEC-field varies as $\delta\theta_{\mathrm{B}}=\theta_{\mathbf{k}}\exp%
\left[ i\left( \mathbf{k}\cdot\mathbf{r}-\omega t\right) \right] $.
The assumption of small amplitude oscillations implies that the
fluctuation amplitudes $\theta_{\mathbf{k}}$ and
$\eta_{\mathbf{k}}\ll1$. Below, we describe their time evolution.

\subsection{BEC dynamics\label{SubSec_BEC}}

In general, the BEC dynamics is conveniently described in terms of
the real
valued density and phase fields, $\rho _{\mathrm{B}}(\mathbf{r};t)$ and $%
\theta _{\mathrm{B}}(\mathbf{r},t)$, the fluctuations of which we
introduced above \cite{PS}. These quantities relate to the complex
valued condensate field as $\phi \left( \mathbf{r},t\right)
=\left\langle \hat{\psi}\left( \mathbf{r},t\right) \right\rangle
=\sqrt{\rho _{\mathrm{B}}\left( \mathbf{r} ,t\right) }e^{i\theta
_{\mathrm{B}}\left( \mathbf{r},t\right) }$. The time-dependent
Gross-Pitaevskii equation (Eq.\ (\ref{tigp}) with $\mu _{
\mathrm{B}}\rightarrow -i\hbar \partial /\partial t$) leads to
\begin{equation}
\frac{\partial \left\vert \phi \right\vert ^{2}}{\partial
t}+\overline{ \nabla }\cdot \left[ \frac{\hbar
}{i2m_{\mathrm{B}}}\left( \phi ^{\ast } \overline{\nabla }\phi -\phi
\overline{\nabla }\phi ^{\ast }\right) \right] =0.  \label{eq:BEC}
\end{equation}
\newline
With the density and phase fields, Eq.\ (\ref{eq:BEC}) takes the
form of a classical continuity equation,
\begin{equation}
\frac{\partial \rho _{\mathrm{B}}(\mathbf{r};t)}{\partial
t}=-\overline{ \nabla }\cdot \left[ \rho
_{\mathrm{B}}(\mathbf{r};t)\mathbf{v}_{\mathrm{s}} \right] \;,
\label{eq:BECdensity}
\end{equation}
where $\mathbf{v}_{\mathrm{s}}=\left( \hbar /m_{\mathrm{B}}\right)
\overline{ \nabla }\theta _{\mathrm{B}}(\mathbf{r};t)$ is a
superfluid velocity. The real part of the Gross-Pitaevskii equation
gives
\begin{align}
-\hbar \frac{\partial \theta _{\mathrm{B}}\left( \mathbf{r},t\right)
}{
\partial t}& =\frac{-\hbar ^{2}}{2m_{\mathrm{B}}\sqrt{\rho _{\mathrm{B}
}\left( \mathbf{r},t\right) }}\overline{\nabla }^{2}\sqrt{\rho
_{\mathrm{B}
}\left( \mathbf{r},t\right) }  \notag \\
& +\lambda _{\mathrm{BB}}\rho _{\mathrm{B}}\left(
\mathbf{r},t\right) +\lambda _{\mathrm{FB}}\rho _{\mathrm{F}}\left(
\mathbf{r},t\right) +\frac{1
}{2}m_{\mathrm{B}}\mathbf{v}_{\mathrm{s}}^{2}(\mathbf{r};t),
\label{eq:BECphase}
\end{align}
which is similar to the Landau superfluid velocity equation with the
Bohm-potential ($\propto \overline{\nabla }^{2}\sqrt{\rho
_{\mathrm{B}}}$) added to the usual mean-field contributions
$\lambda _{\mathrm{BB}}\rho _{ \mathrm{B}}+\lambda
_{\mathrm{FB}}\rho _{\mathrm{F}}$ and the superfluid kinetic energy,
$m_{\mathrm{B}}\mathbf{v}_{\mathrm{s}}^{2}/2$.

Linearizing the BEC-fields as in Sec.\ \ref{SubSec_fermion}, then
substituting them into Eqs.\ (\ref{eq:BECdensity}) and
(\ref{eq:BECphase}) give
\begin{align}
-i\omega\eta_{\mathbf{k}} & =\frac{\hbar
k^{2}}{m_{\mathrm{B}}}\theta_{
\mathbf{k}}, \\
i\hbar\omega\theta_{\mathbf{k}} & =\left(
\frac{\hbar^{2}k^{2}}{4m_{\mathrm{B
}}}+\lambda_{\mathrm{BB}}\rho_{\mathrm{B}}^{0}\right)
\eta_{\mathbf{k} }+\lambda_{\mathrm{FB}}\delta\rho_{F,\mathbf{k}}\;,
\end{align}
where $\delta\rho_{\mathrm{F},\mathbf{k}}$ is the Fourier transform
of the fermion density fluctuation that corresponds to
Eq.(\ref{an}). The fluctuation amplitude of the boson density is
then proportional to the fermion density fluctuation and can be
written as
\begin{equation}
\eta_{\mathbf{k}}=\frac{\lambda_{\mathrm{FB}}}{\lambda_{\mathrm{BB}}}\frac{
\delta\rho_{\mathrm{F},\mathbf{k}}/\rho_{\mathrm{B}}^{0}}{[\omega
/kc]^{2}-[1+(k\xi)^{2}]}\;,   \label{eq:eta}
\end{equation}
where $c$ is the sound velocity of the pure BEC, $c=\left( \lambda
_{\mathrm{ BB}}\rho_{\mathrm{B}}^{0}/m_{\mathrm{B}}\right) ^{1/2}$.
Note that the sign of the ratio of the boson to fermion density
fluctuations depends on the frequency $\omega$.

\subsection{Dispersion relation of the collective modes}

Returning to the Fermi-liquid dynamics of Sec.\ \ref{SubSec_BEC}, we
write Eq.\ (\ref{eq:fermiondynamics}) as
\begin{equation}
\left[ \frac{\omega}{kv_{\mathrm{F}}}-\mathbf{\cos}\left(
\varphi\right) \right]
u_{\mathbf{p}}-\eta_{\mathbf{k}}\;\mathbf{\cos}\left( \varphi
\right) \left[
\frac{\lambda_{\mathrm{FB}}\rho_{\mathrm{B}}^{0}}{2\epsilon_{
\mathrm{F}}}\right] =0,   \label{fermion dynamics}
\end{equation}
where $\varphi$ is the angle between the $\mathbf{k}$ and
$\mathbf{p}$ --vectors, and where $\epsilon_{\mathrm{F}}$ represents
the equilibrium Fermi-energy.

From the definition of $n$, which implies $\rho
_{F}({\mathbf{r}},t)=\hbar^{-3}\int d^{3}p$
$n(\mathbf{r},\mathbf{p},t)$, and with Eq.\ (\ref{an}), we find that
the Fermi-density fluctuation $\delta \rho _{F,\mathbf{k}}$ in Eq.\
(\ref{eq:eta}) is proportional to the angular average of the
deformation amplitude,
\begin{equation}
\delta \rho _{F,\mathbf{k}}=3\rho _{F}^{0}\langle u\rangle \;,
\label{drho}
\end{equation}%
where $\langle u\rangle =(4\pi )^{-1}\int d\Omega _{{\mathbf{p}}}u_{{\mathbf{%
p}}}$. Combining Eqs.\ (\ref{eq:eta}) and (\ref{fermion dynamics}),
we obtain
\begin{equation}
\left[ \frac{\omega }{kv_{\mathrm{F}}}-\mathbf{\cos }\left( \varphi \right) %
\right] u_{\mathbf{p}}=\mathbf{\cos }\left( \varphi \right)
\frac{F}{[\omega /kc]^{2}-[1+(k\xi )^{2}]}\langle u\rangle \;,
\label{collective dynamics}
\end{equation}%
where $F=(\lambda_{\mathrm{FB}}/\lambda_{\mathrm{BB}}) (3\lambda
_{\mathrm{FB}}\rho _{F}^{0}/2\epsilon _{\mathrm{F}})$ is the
interaction parameter of the system that embodies all interaction
information necessary for understanding the zero-sound mode
behavior. Alternatively, we can write
\begin{equation}
F=\frac{k_{\mathrm{F}}a_{\mathrm{FB}}}{2\pi
}\frac{a_{\mathrm{FB}}}{a_{ \mathrm{B}}}\left(
1+\frac{m_{\mathrm{F}}}{m_{\mathrm{B}}}\right) \left( 1+
\frac{m_{\mathrm{B}}}{m_{\mathrm{F}}}\right) ,
\end{equation}
to expresses the interaction dependence in terms of the scattering
lengths.

We find that the angular dependence of the mode fluctuation is
restricted by the density-density coupling of the liquids in the
dilute fermion-BEC
mixture. Expressing $u_{\mathbf{p}}$ in a spherical harmonics form, $u_{%
\mathbf{p}}=\sum_{m,l}u_{lm}Y_{lm}$, we obtain that $u_{lm}=0$ for
$m\neq0$. In the collective excitation, the local Fermi momentum
distribution fluctuates as a longitudinal mode with an azimuthal
symmetry around the direction of propagation with the deformation
amplitude $u_{\mathbf{p}}$, so that the sole momentum dependence is
upon the angle between $\mathbf{k}$ and $\mathbf{p}$. Rather than
writing $u_{\mathbf{p}}=\sum_{l}u_{l0}Y_{l0}$, we simply determine
the dependence upon $\varphi$ by solving for $u_{\mathbf{p}} $ from
Eq.(\ref{collective dynamics}) and take the angular average.
Defining the scaled phase velocity as
$s\equiv\omega/kv_{\mathrm{F}}$, we obtain
\begin{equation}
\frac{s^{2}\left( v_{\mathrm{F}}/c\right) ^{2}-\left[ 1+\left(
k\xi\right) ^{2}\right]
}{F}=\frac{1}{2}\int_{-1}^{1}\frac{x}{s^{\ast}-x}dx\;,
\label{eq:collective dynamics}
\end{equation}
where we set $x\equiv\cos\varphi$. From Eq.\ (\ref{eq:collective
dynamics} we can determine the dispersion relation of the collective
modes. The complex conjugate, $s^{\ast}$, follows from the
appropriate choice of boundary conditions as we explain in the next
section.

The structure of the excited state is apparent from neither Eq.\ (\ref%
{eq:collective dynamics}) nor Landau's semiclassical description (see \cite%
{Goldstone}). We believe that the actual excited state can be
written a linear superposition of two states: One is a product state
of the ground state BEC and an excited fermion state, which is
itself a linear superposition of particle-hole excitations as is the
case in the pure fermi-liquid zero-sound mode. The other is the
product of the ground state of the Fermi-liquid and a BEC-state that
contains a single Bogoliubov quasi-particle excitation.

\section{Mode analysis}

Equation (\ref{eq:collective dynamics}) is a close relative of the
one that describes the zero sound excitation of a two-component pure
Fermi-liquid whose inter-particle interactions are restricted to
s-wave partial waves \cite{Baym}
\begin{equation}
\frac{1}{F_{0}}=-1+\frac{s}{2}\ln\left( \frac{s+1}{s-1}\right) ,
\label{eq:pureFermiliquid}
\end{equation}
where $F_{0}$ is the Landau interaction parameter.

When $s>1$, the solution $s$ to Eq.\ (\ref{eq:collective dynamics})
is real and the right-hand side of Eq.\ (\ref{eq:collective
dynamics}) reduces to a simple expression
\begin{equation}
\frac{\left[ s^{2}(v_{\mathrm{F}}/c)^{2}-\left( 1+(k\xi)^{2}\right)
\right] }{F}=-1+\frac{s}{2}\ln\left( \frac{s+1}{s-1}\right) \;,
\label{eq:s>1dispersion}
\end{equation}
of the form of Eq.\ (\ref{eq:pureFermiliquid}). The similarity
suggests identifying $F/\left[ s^{2}(v_{\mathrm{F}}/c)^{2}-\left(
1+(k\xi )^{2}\right) \right] $ as an effective Landau-liquid
parameter. However, a true Landau-liquid parameter does not depend
on the wavenumber $k$. On the other hand, the appearance of the
wavenumber leads to the rich mode structure that we discuss in the
following subsections.

As mentioned before, $F$ parameterizes all the dependence on the
interaction
parameters, and zero-sound modes exist for $0<F<1$. The eigenvalues of Eq.\ (%
\ref{eq:pureFermiliquid}) depend on the velocity ratio
$v_{\mathrm{F}}/c$, which quantifies the ratio of the time scales of
the boson and fermion dynamics. In fact, in the long wavelength
limit, the oscillation periods of both the BEC and fermion modes of
wavenumber $k$ are order of $(kc)^{-1}$ and
$(kv_{\mathrm{F}})^{-1}$. These time-scales are also associated with
the response time to density perturbations of spatial variation of
$k^{-1}$.

\subsection{Collective mode excitations $0<F<1$}

If $v_{\mathrm{F}}\gg c$ and $0<F<1$, two collective modes exist: an
undamped mode with a real $s$ and a damped mode with a complex $s$.
The imaginary part\ of $s$ describes the decay rate of the damped
mode and the zero-sound analogy reveals that the decay mechanism is
indeed Landau damping: the creation of particle-hole pairs in the
Fermi-system. Landau damping only occurs when $s<1$ due to energy
conservation: a particle-hole
excitation with momenta $\mathbf{K}+\mathbf{k}/2$ (particle) and $\mathbf{K}-%
\mathbf{k}/2$ (hole) near the Fermi-surface of momentum $\mathbf{K}$
has an
oscillation frequency that is smaller than $\omega$. In other words, $\hbar%
\mathbf{K}\cdot\mathbf{k}/m_{F}\,<v_{F}k$, so that a single particle
hole excitation cannot absorb all the energy of the collective
excitation.

\subsubsection{Undamped mode}

\begin{figure}[ptb]
\begin{center}
\includegraphics[width=3.1in]{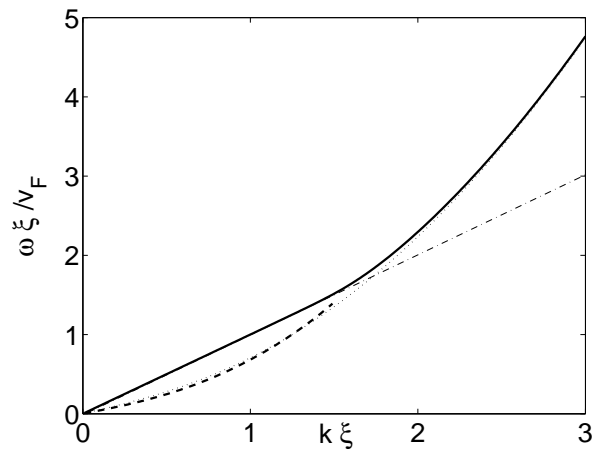}
\end{center}
\caption{The dispersion relation ($\protect\omega\protect\xi/v_{F}$
vs $k \protect\xi$) for a fixed $F=0.5$ and a fixed $v_{F}/c=2$. The
dispersion changes from linear to Bogoliubov dispersion.}
\label{fig:dispfp5a2}
\end{figure}

\begin{figure}[ptb]
\begin{center}
\includegraphics[width=3.1in]{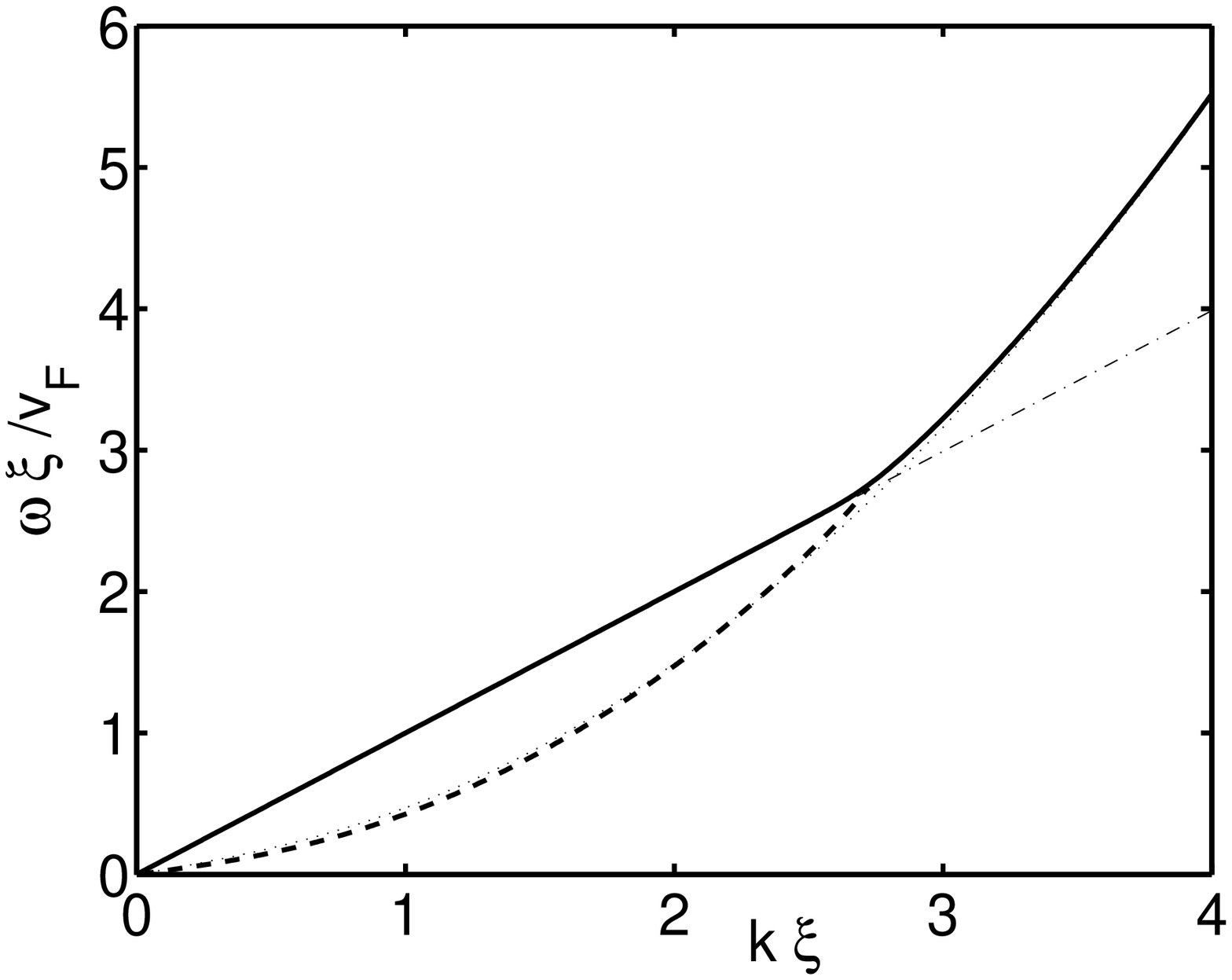}
\end{center}
\caption{The dispersion relation ($\protect\omega\protect\xi/v_{F}$
vs $k \protect\xi$) for a fixed $F=0.5$ and a fixed $v_{F}/c=3$. The
dispersion changes from linear to Bogoliubov dispersion.}
\label{fig:dispfp5a3}
\end{figure}

\begin{figure}[ptb]
\begin{center}
\includegraphics[width=3.1in]{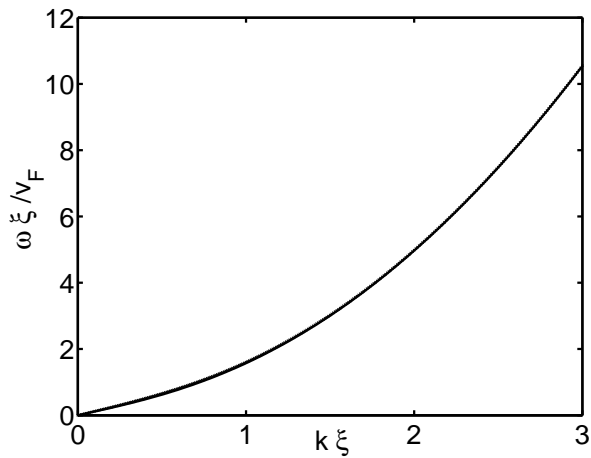}
\end{center}
\caption{The dispersion relation ($\protect\omega\protect\xi/v_{F}$
vs $k \protect\xi$) for a fixed $F=0.5$ and a fixed $v_{F}/c=0.9$.
The zero sound dispersion follows the Bogoliubov dispersion for all
$k$ values.} \label{fig:dispfp5ap9}
\end{figure}

The scaled dispersion relations of the undamped modes, $\omega\xi /v_{%
\mathrm{F}}$ vs. $k\xi$, at various velocity ratios of
$v_{\mathrm{F}}/c$
with a fixed $F=0.5$ are shown in Figs.\ \ref{fig:dispfp5a2}, \ref%
{fig:dispfp5a3}, and \ref{fig:dispfp5ap9}. For $v_{\mathrm{F}}\geq
c$ the dispersion is linear with the phase velocity $v\approx
v_{\mathrm{F}}$ at small $k$, similar to the long wavelength
dispersion of the zero-sound mode in a weakly interacting pure
Fermi-liquid. At larger $k$, the dispersion becomes identical to
that of an elementary excitation of a single
homogeneous BEC (i.e., Bogoliubove mode) with $\omega=kc\sqrt{1+(k\xi)^{2}}$%
. The transition from linear dispersion to Bogoliubov dispersion
occurs gradually around
$k_{\mathrm{trans}}=\xi^{-1}\sqrt{(v_{F}/c)^{2}-1}$ as a
result of the avoided crossing of the two dispersions (See Figs.\ \ref%
{fig:dispfp5a2} and \ref{fig:dispfp5a3}.) On the other hand, when $v_{%
\mathrm{F}}<c$, the undamped mode dispersion is the Bogoliubov
dispersion for the entire range of $k$-values, as seen in Fig.\
\ref{fig:dispfp5ap9}.

Since our treatment of the Fermi-liquid dynamics assumes $k\ll
k_{\mathrm{F}} $, we need to check the wavenumber regime of
validity. However, for mixtures that consist of light boson and
heavy fermion atoms (for example, $^{1}$H and $^{40}$K, or $^{7}$Li
and $^{87}$Rb), this range can extend to
wavenumbers comparable to or greater than $\xi^{-1}$, as implied by $k_{%
\mathrm{F}}\xi=(m_{\mathrm{F}}/2m_{\mathrm{B}})(v_{\mathrm{F}}/c)$.

If $v_{F}<c$, the BEC responds sufficiently fast to follow the
fermion oscillations adiabatically and the static description of
BEC-mediated interaction is valid. In this case, the BEC-mediated
interaction is attractive and there is no Fermi-liquid like
zero-sound mode. Instead, there is a collective oscillation that
resembles the elementary excitation of a pure BEC with $v\simeq c$
for all range of $k$ (Fig. \ref{fig:dispfp5ap9}). On the other hand,
if $v_{F}\gg c$, the BEC response is significantly slower than the
fermion frequency and the BEC oscillates out of phase with the
fermions. The overall sign of the left hand side in Eq.\
(\ref{eq:collective dynamics}) is positive for small $k$, giving an
effective repulsive fermion-fermion interaction. In this regime, the
collective fermion mode with zero-sound characteristics occurs with
phase velocity velocity $ v\approx v_{\mathrm{F}}$ (see Figs.
\ref{fig:dispfp5a2} and \ref{fig:dispfp5a3}).

\subsubsection{Damped mode}

In addition to the undamped mode, a damped collective excitation
exists in $0<F<1$. As stated before, Landau damping \cite{PN} is the
dominant damping mechanism of zero-sound in the regime of
collisionless dynamics.

The imaginary part of the complex eigenvalues for Eq.\
(\ref{eq:collective dynamics}) is proportional to the damping rate.
Assigning the correct sign to the imaginary part of $s$ in the
eigenvalue equation Eq.\ ( \ref{eq:collective dynamics}) is rather
delicate: While the damped mode corresponds to $s=r-i\gamma$ with
$r,\gamma>0$, causality requires that the integral on the right-hand
side, which describes the retarded fermion-mediated boson-boson
interaction, acquires $s^{\ast}=s+i\gamma$. The damped excitation
evolves in time as $\exp\left[ -i\left( \omega -i\Gamma/2\right)
t\right] $ with $\Gamma=2kv_{\mathrm{F}}\gamma$. Independent from
the collective mode equation, we estimate the damping rate $
\Gamma(k)$ using the Fermi-golden rule. The initial state
$|i\rangle$ is the product of the single quasi-particle BEC state
with momentum $\mathbf{k}$ and a fermion state that consists of a
completely filled Fermi sphere. The final state $|f\rangle$ is the
product of the BEC-ground state and an excited fermion state with a
particle-hole pair of momenta $\mathbf{K}+ \mathbf{k}/2$ for the
particle and $\mathbf{K}-\mathbf{k}/2$ for the hole. The
quasi-particle-hole pair is created near the Fermi-surface of
momentum $\mathbf{K}$. Treating the fermion-boson density-density
interaction as a perturbation $\hat{H}_{p}$ and summing over the
final fermion states, we obtain the damping rate
\begin{equation}
\Gamma_{\mathrm{FG}}(k)=\left( 2\pi/\hbar\right) \sum_{f}|\langle f|\hat {H}%
_{p}|i\rangle|^{2}\delta\left( E_{f}-E_{i}\right) =\frac{\pi
F}{2}\left( \frac{c}{v_{\mathrm{F}}}\right) kc\;,   \label{eq:F-G
rule}
\end{equation}
where the energy $E_{i}$ and $E_{f}$ represent the energies of the
initial and final state, respectively. The damping rate is
proportional to $kc$ as well as $F$ and $\left(
v_{\mathrm{F}}/c\right) ^{-1}$.

\begin{figure}
[ptb]
\begin{center}
\includegraphics[width=3.1in]{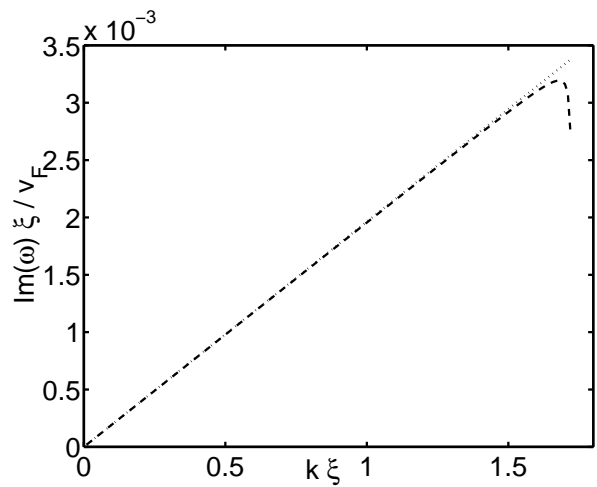}
\caption{$v_{\mathrm{F}}/c=2$, $F=0.01$. The dashed line shows the
numerically evaluated $\operatorname{Im}[\omega\xi/v_{\mathrm{F}}]$,
while the dotted line plots $\operatorname{Im}
[\omega\xi/v_{\mathrm{F}}]$ evaluated from the Fermi-Golden Rule.}
\label{fig:dispfgrule-numfp01}
\end{center}
\end{figure}

\begin{figure}
[ptb]
\begin{center}
\includegraphics[width=3.1in]{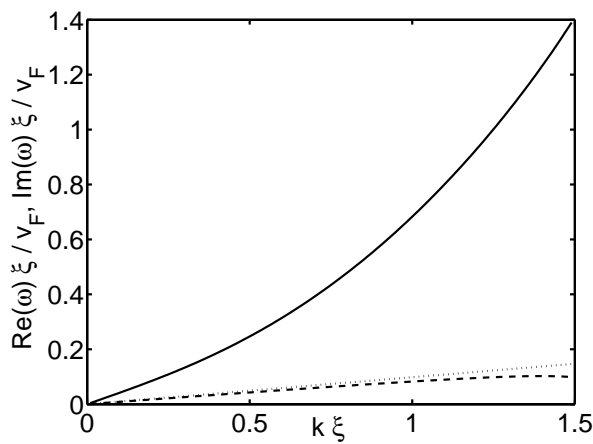}
\caption{$v_{\mathrm{F}}/c=2$, $F=0.5$. The solid line shows the
numerically evaluated values of
$\operatorname{Re}[\omega\xi/v_{\mathrm{F}}]$, the dashed line plots
the numerically evaluated curve of
$\operatorname{Im}[\omega\xi/v_{\mathrm{F}}]$, whereas the dotted
line gives the values of
$\operatorname{Im}[\omega\xi/v_{\mathrm{F}}]$ determined from the
Fermi-Golden Rule calculation.} \label{fig:dispfgrule-numfp5}
\end{center}
\end{figure}

Note that Eq.\ (\ref{eq:F-G rule}) is valid for $k\ll
k_{\mathrm{F}}$. Figures \ref{fig:dispfgrule-numfp01} and
\ref{fig:dispfgrule-numfp5} show
$\operatorname{Im}[\omega]\xi/v_{\mathrm{F}}$ obtained both by
numerics from Eq.\ (\ref{eq:collective dynamics}) and by the
Fermi-golden rule estimate for $v_{\mathrm{F}}/c=2$, and $F=0.01$
and $F=0.5$, respectively. For $0<F\ll1$ with $v_{\mathrm{F}}/c>1$,
the rate of damping is orders of magnitude smaller than the
excitation frequency, the phase velocity of the damped mode is
almost equal to that of the BEC sound mode, and the Fermi-Golden
rule approximation is in good agreement with the numerical solutions
to Eq.\ (\ref{eq:collective dynamics}).

The damped mode exists for small wavenumbers $k$ and ends at the
`termination point', $k_{\mathrm{end}}$, as seen in Figs.\ \ref
{fig:dispfgrule-numfp01} and \ref{fig:dispfgrule-numfp5}. This
behavior is reminiscent of the excitations in $^{4}$He
\cite{Landau}. In the wavenumber interval that the mode exists, the
rate for Landau-damping does not increase monotonically with $k$.
This behavior is somewhat counter-intuitive as an increase in $k$
generally increases the phase space volume of the available final
states. The termination point $k_{\mathrm{end}}\xi$ expressed in
terms of $\gamma$ is
\begin{equation}
k_{\mathrm{end}}\xi=\left( 1-\gamma^{2}\right) ^{1/2}\left\{ \left(
\frac{v_{ \mathrm{F}}}{c}\right) ^{2}-1+F+\frac{F}{4}\left[
\pi\gamma -\frac{ 3\gamma^{2}}{4}-2\ln\left( \frac{\gamma}{2}\right)
\right] \right\} ^{1/2}\;,
\end{equation}
which can be tested experimentally.

The magnitude of $\operatorname{Im}[\omega]$ is significantly
smaller than $\operatorname{Re}[\omega]$  even for a system of
relatively strong interactions such as $F=0.5$. Since we expect the
collective excitations to be good quasi-particles, if the lifetime
of the mode is short enough, the damping of the mode can be measured
in atom trap experiments. The time scale of the mode damping is set
by
\begin{equation}
\tau_{0}=\frac{\xi}{2v_{\mathrm{F}}}=\frac{m_{\mathrm{F}}}{m_{\mathrm{B}}}
\frac{1}{ck_{\mathrm{F}}}\;.   \label{tau0}
\end{equation}
$\tau_{0}$ is the time that takes a particle with velocity $v_{F}$
to travel half the distance of the BEC healing length.
Alternatively, we write $k_{ \mathrm{F}}$ in terms of the average
inter-fermion distance as $r_{\mathrm{F} }=\left[
\rho_{\mathrm{F}}^{0}\right] ^{-1/3}$, $k_{\mathrm{F}}=\left(
6\pi^{2}\right)^{1/3} \left[\rho_{\mathrm{F}}^{0}\right]^{1/3}\simeq
3.9/r_{\mathrm{F}}$, and find that $\tau_{0}$ is proportional to the
time a particle requires to travel at the velocity of the BEC-sound
in order to cover the average inter-fermion distance
$\tau_{0}\approx(m_{\mathrm{F}}/3.9m_{\mathrm{B}
})\times(r_{\mathrm{F}}/c)$. Under typical experimental conditions
$\tau_{0}$ can be a few milliseconds, which is sufficiently long to
measure the damping and sufficiently short to keep a propagating
wavepacket from leaving the trap region where the system is
approximately homogeneous.

Since the damped mode can be viewed as an elementary BEC-excitation
that has been modified by the fermion-mediated boson-boson
interactions, the decay can also occur by Belyaev damping where
pairs of boson quasi-particles are created \cite{Belyaev}. While our
treatment does not describe this damping, the decay rate due to
Belyaev damping tends to be slow in the long wavelength limit $k<\xi
^{-1}$, $\Gamma _{\mathrm{Bel}}\approx (6\sqrt{\pi } /5)\sqrt{\rho
_{B}^{0}a_{BB}^{3}}(k\xi )^{4}ck$, thus can be neglected in the
range of $k$ of our interest.

\begin{figure}[ptb]
\begin{center}
\includegraphics[width=3.1in]{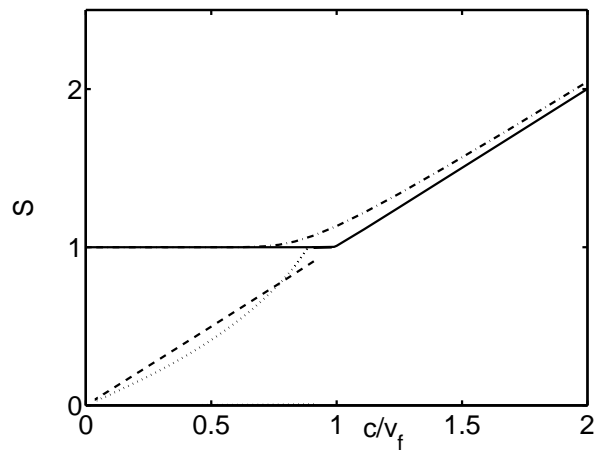}
\end{center}
\caption{Plot showing the solution of Eq.\
(\protect\ref{eq:collective dynamics}) for $s$ at $k \protect\xi=0$
as a function of $c/v_{F}$ . The solid line shows the undamped mode
phase velocity with $F=0.01$, the dash-dotted line plots its value
for the undamped mode with $F=0.5$, whereas the dashed line shows
the values for the damped mode with $F=0.5$ and the dotted line for
the damped mode with $F=0.01$.} \label{fig:fp01fp5mode12}
\end{figure}

Finally, it is instructive to express $s$ as a function of the
velocity ratio, $v_{\mathrm{F}}/c$, in the limit $k\rightarrow0$
(see Fig.\ \ref {fig:fp01fp5mode12}.)

Similar to the behavior of the dispersion relations, there is a
transition of the long wavelength phase velocity $v$, from its
undamped mode $v\approx v_{F}$ to the BEC-sound velocity $v\approx
c$. The smoothness of this transition depends on the value of $F$:
the smaller the value of $F$ the
more abrupt the transition occurs with varying velocity ratio. If $v_{%
\mathrm{F}}<c$, the static description of the BEC-mediated
fermion-fermion interaction is valid and the fermions experience an
effective mutual attraction that does not support a long lived
zero-sound excitation in a two-component pure Fermi-liquid system.
We can determine the transition
point (or region), $\left( v_{\mathrm{F}}/c\right) _{\mathrm{trans}}$. For $%
k\rightarrow0$, the BEC and zero sound modes mix strongly near $v_{\mathrm{F}%
}/c\simeq1$, similar to the region where the linear dispersion
merges into a Bogoliubov dispersion as we have seen in the
dispersion relations. Quantifying the deviation of $s$ from unity by
setting $s=1+\epsilon$, where $0<\epsilon\ll1$ in the
Eq.(\ref{eq:collective dynamics}) we find
\begin{equation}
\left. \frac{v_{\mathrm{F}}}{c}\right\vert
_{\mathrm{trans}}\simeq\left[ 1+F\left(
\frac{1}{2}\ln\frac{2}{\epsilon}-1\right) \right] ^{-1/2}\;.
\label{transitionpoint v/c}
\end{equation}
The avoided crossing, termination point and strong mode mixing occur
in the velocity ratio interval between $\left(
v_{\mathrm{F}}/c\right) _{\mathrm{trans}}$ and $1$.

\subsection{Instability and phase transition}

Equation (\ref{eq:collective dynamics}) provides insight into the
stability, or lack thereof, of a homogeneous fermion-boson mixture.
If $F>1$, $s$ the eigenfrequency of the damped modes of longest
wavelengths take on purely imaginary values. The purely imaginary
value, and not just the acquisition of an imaginary part as in
\cite{PZWM02} (a claim that was also corrected in \cite{Yip2}),
signals the instability of the homogeneous mixture. Then Eq.\ (
\ref{eq:collective dynamics}) is reduced to
\begin{equation}
\frac{\gamma ^{2}\left( v_{\mathrm{F}}/c\right) ^{2}+\left[ 1+\left(
k\xi \right) ^{2}\right] }{F}=1+\gamma \arctan \left(
\frac{1}{\gamma }\right) \;. \label{eq:instability dispersion}
\end{equation}
Since the unstable mode is the lowest frequency mode, i.e., the
damped or BEC-sound like mode in the stable mixture, we can say that
it is the fermion-mediated boson-boson interaction that triggers
triggers the instability of the homogeneous BEC in the mixture.
Since $F$ depends on $k_{\mathrm{F}}$, the instability condition $
F>1$ can be expressed as a condition on the fermion density: $\rho _{\mathrm{%
\ F}}^{0}>\rho _{\mathrm{F},\mathrm{crit}}$, where
\begin{equation}
\rho _{\mathrm{F},\mathrm{crit}}=\frac{4\pi
}{3a_{\mathrm{FB}}^{3}}\left[
\frac{a_{\mathrm{BB}}/a_{\mathrm{FB}}}{(1+m_{\mathrm{B}}/m_{\mathrm{F}%
})(1+m_{\mathrm{F}}/m_{\mathrm{B}})}\right] ^{3}\;.
\label{eq:rhocrit}
\end{equation}%
If the equilibrium fermion density exceeds $\rho _{\mathrm{F},\mathrm{crit}}$%
, the Fermi-liquid becomes immiscible to the BEC and spontaneously
breaks the translational symmetry by undergoing phase separation
into either (a) pure fermion and pure BEC or (b) pure fermion and
mixed fermion-BEC phases \cite{Viverit}.

Equation\ (\ref{eq:instability dispersion}) also provides insight
into the dynamics of the onset (the early stages) of the phase
separation instability. Consider an experiment that brings
$\rho_{\mathrm{\ F}}$ in a fermion-BEC mixture to above
$\rho_{\mathrm{F},\mathrm{crit}}$ by an abrupt change of
$a_{\mathrm{BB}}$ (\ref{eq:rhocrit}). In response, the amplitude of
the longest wavelength modes of the damped zero-sound excitations
grow exponentially at a rate $\gamma(k)k\xi/\tau_{0}$ nucleating
spatial regions of pure fermion matter when the dynamics becomes
non-linear. Those modes that grow fastest (with wavenumber $k_{d}$)
dominate the dynamics and we expect them to determine the size of
the single phase matter clusters and the rate of the phase
separation \cite{BECphase}.

When $\gamma$ is sufficiently small, we can expand the right-hand
side of Eq.\ (\ref{eq:instability dispersion}) and determine $k_{d}$
analytically. The modes with wavenumber in the range, $k\xi\in\left[
0,\sqrt{F-1}\right] $ are unstable and grow exponentially at the
rate $R=\gamma k\xi\tau_{0}^{-1}$ with
\begin{equation}
R\tau_{0}=\left\{ \frac{\pi F}{4}+\sqrt{\left( \frac{\pi
F}{4}\right)
^{2}-\left( \frac{v_{\mathrm{F}}}{c}\right) ^{2}\left( 1-F+q^{2}\right) }%
\right\} k\xi\;.
\end{equation}
The maximum growth rate occurs at $k_{d}$, where
\begin{equation}
\left( k_{d}\xi\right) {}^{2}=\frac{F-1}{2}+\frac{\pi F}{128}\left( \frac{v_{%
\mathrm{F}}}{c}\right) ^{-2}\left. \left[ 3\pi F-\sqrt{(3\pi
F)^{2}+128\left( \frac{v_{\mathrm{F}}}{c}\right) ^{2}\left( F-1\right) }%
\right] \right\} ,
\end{equation}
with the average cluster size $\sim2\pi/k_{d}$.

\section{Experimental aspects\label{Sec_expt}}

In this section, we mention relevant atom trap techniques and
estimate magnitudes of quantities to gauge the feasibility of
experimentally observing the collective mode physics. The important
parameter for the boson-boson interactions in a dilute BEC is the
dimensionless gas parameter, $D=\sqrt
{\rho_{\mathrm{B}}^{0}a_{\mathrm{BB}}^{3}}$ \cite{LHY}. We find it
useful to express the density in units of a reference density $\rho_{\mathrm{%
R}}$ that corresponds to the lower range obtained in atom traps, $\rho_{%
\mathrm{R}}=10^{12}\mathrm{cm}^{-3}$. Then, the gas parameter is
\begin{equation}
D=\sqrt{\frac{\rho_{\mathrm{B}}}{\rho_{\mathrm{R}}}\frac{a_{\mathrm{BB}}}{1%
\mathrm{nm}}}\;3.2\times10^{-5}\;.   \label{d}
\end{equation}
We introduce the atomic mass number of the bosonic and fermionic isotopes, $%
A_{\mathrm{B}}$ and $A_{\mathrm{F}}$ ($A_{\mathrm{F}}=6$ for
$^{6}$Li, for instance) and express
$\lambda_{\mathrm{BB}}\rho_{\mathrm{B}}^{0}$ and Fermi-energies
$\epsilon_{F}$ as
\begin{align}
\lambda_{\mathrm{BB}}\rho_{\mathrm{B}}^{0} & =\frac{6\mathrm{nK}}{A_{\mathrm{%
B}}}\frac{\rho_{\mathrm{B}}^{0}}{\rho_{\mathrm{R}}}\frac{a_{\mathrm{BB}}}{1%
\mathrm{nm}}\;,  \notag \\
\epsilon_{\mathrm{F}} & =\frac{3.65\mathrm{\mu
K}}{A_{\mathrm{F}}}\left(
\frac{\rho_{\mathrm{F}}}{\rho_{\mathrm{R}}}\right) ^{2/3}\;.
\label{om}
\end{align}
The time scale for the fermion dynamics is equal to
\begin{equation}
t_{\mathrm{F}}=\frac{\hbar}{\epsilon_{\mathrm{F}}}\approx2.09\mu \mathrm{sec}%
\times A_{\mathrm{F}}\left( \frac{\rho_{\mathrm{F}}^{0}}{\rho_{\mathrm{R}}}%
\right) ^{-2/3}\;.,   \label{tf}
\end{equation}
from which it follows that the Fermi velocity can be of the order of
several cm/sec,
\begin{equation}
v_{\mathrm{F}}=\frac{2}{t_{\mathrm{F}}k_{\mathrm{F}}}\approx\frac {24.5}{A_{%
\mathrm{F}}}\mathrm{cm/sec}\times\left( \frac{\rho_{\mathrm{F}}^{0}}{\rho_{%
\mathrm{R}}}\right) ^{1/3}\;.   \label{vf}
\end{equation}
Most importantly, the velocity ratio $v_{\mathrm{F}}/c$ which
quantifies the ratio of the relevant fermion to boson time scales
and which determines the validity of the static approximation of the
BEC-mediated interaction, can be written as
\begin{equation}
\frac{v_{\mathrm{F}}}{c}=\frac{A_{\mathrm{B}}}{A_{\mathrm{F}}}\frac
{0.55}{ D^{1/3}}\left(
\frac{\rho_{\mathrm{F}}^{0}}{\rho_{\mathrm{B}}^{0}}\right) ^{1/3}\;.
\label{vfoc}
\end{equation}
Clearly experiments can access both the $v_{\mathrm{F}}/c<1$ and
$v_{\mathrm{F}}/c>1$ regimes. For example, $^{6}$Li with
$\rho_{\mathrm{F}}^{0}=10^{13} \mathrm{cm}^{-3}$ immersed in
$^{23}$Na BEC with $\rho _{\mathrm{B}
}^{0}=3\times10^{14}\mathrm{cm}^{-3}$ and
$a_{\mathrm{BB}}\sim1.5\mathrm{nm}$ gives
$v_{\mathrm{F}}/c\approx0.46$, whereas changing the density of the
BEC to $\rho_{\mathrm{B}}^{0}=3\times10^{12}\mathrm{cm}^{-3}$
increase the velocity ratio to $v_{\mathrm{F}}/c\approx4.6$.

To observe the predicted collective modes an experimentalist could
measure the group velocities in a fermion-boson mixture that is
contained in a cigar shaped trap. We assume that the transverse
confinement is not so tight as to bring the system into the regime
where it becomes effectively one-dimensional. A sudden perturbation
near the trap middle created by a focused laser beam can excite both
collective modes. In response, two wavepackets of different
densities propagate outward with different group velocities. The
density variations (fermion and boson) can be imaged directly, as in
the first observation of the BEC-sound mode \cite{Ketterlesound}.
The slower wavepacket would correspond to the Landau damped mode
with
\begin{align}
\frac{1}{\tau_{0}} & \approx\frac{110}{A_{\mathrm{F}}}\mathrm{{msec}
^{-1}\;\times}  \notag \\
& \;\;\;\;\;\;\;\;\left(
\frac{\rho_{\mathrm{F}}^{0}}{\rho_{\mathrm{R}}} \right)
^{1/3}\sqrt{\left( \frac{a_{\mathrm{BB}}}{1\mathrm{nm}}\right)
\;\left( \frac{\rho_{\mathrm{B}}^{0}}{\rho_{\mathrm{R}}}\right) }\;,
\label{tau0inverese}
\end{align}
which can be slow enough to observe significant propagation while
sufficiently fast to measure damping before the wavepacket reaches
equilibrium densities that are significantly different from trap
middle.

In more sophisticated experiments, the initial perturbation can be
controlled, and an analysis of the time evolution of the shape of
the propagating wavepackets can be used to infer the dispersion
relations and damping rates. In the past, the greatest control was
achieved by accessing a two-photon resonance to cause low intensity
Bragg scattering. In such an experiment the momentum of the
excitation can be varied by changing the angle of the crossed laser
beams. The laser beams need to be focused near the middle of the
trap to probe the nearly uniform region of the fermion-boson
mixture. Similar techniques have been used in 'Bogoliubov
spectroscopy experiments' \cite{spectroscopy}.

Finally, we remark that by tuning the boson-boson scattering length
close to its zero-point $a_{\mathrm{BB}}\approx 0$, thereby lowers
the critical fermion density can always trigger the phase separation
phenomenon. By triggering the phase separation in a cigar shaped
trap, the surface tension can prevent the single phase domains to
move past each other, the experimenter can obtain a string of single
phase `droplets'.  The size of these domains reflect the dominant
mode wave vector.  Such experimental study was demonstrated in the
phase separation of multi-component BEC's \cite{Ketterleps}.

\section{Conclusions}

We have analyzed the collective modes of a single component
fermion--BEC mixture in ultra-low temperature where the fermions
behave as normal Fermi-liquid.

If $v_{\mathrm{F}}>c$ and $0<F<1$, the homogeneous mixture is
mechanically stable and two collective modes with zero-sound
character exist. One is an undamped mode that resembles the
zero-sound mode of a pure Fermi-liquid system. The dispersion of
this mode is linear with phase velocity $\omega /k\approx
v_{\mathrm{F}}$ in the long wavelength limit and then merges into
the Bogoliubov dispersion of an elementary BEC excitation near $k_{\mathrm{%
trans}}=\xi ^{-1}\sqrt{(v_{\mathrm{F}}/c)^{2}-1}$. The other mode
resembles the sound mode of a pure BEC system in the long wavelength
limit. The mode undergoes Landau damping and decays at the rate
$\Gamma (k)=\gamma (k)k\xi /\tau _{0}$. The dispersion of this mode
terminates at a well-defined wavenumber $k_{\mathrm{end}}$. In the
long wavelength limit, the rate for
Landau damping can be approximated using the Fermi-golden rule and is $%
\Gamma _{\mathrm{FG}}(k)=\left( \pi F/2\right) \left( c/v_{\mathrm{F}%
}\right) ck$.

On the other hand, if $v_{\mathrm{F}}<c$ and $0<F<1$, the BEC
response is sufficiently fast to follow the fermion oscillations
adiabatically, thus the dispersion is the Bogoliubov dispersion for
all $k$.

When $F>1$, we have $\rho_{\mathrm{\
F}}^{0}>\rho_{\mathrm{F},\mathrm{crit}}$ the homogenous mixture is
unstable and undergoes phase separation by forming
clusters of pure fermions. Our analysis shows that the growth rate is $%
R=\gamma k\xi\tau_{0}^{-1}$\ and the average cluster size is $\sim2\pi/k_{d}$%
.

Finally, our discussion of current atom trap technology and relevant
observables indicate that the collective modes can be observed and
the dispersions and damping rates may be measured. In addition, the
phase separation can be triggered, enabling the phase separation
dynamics characterization as soon as the fermions are cooled into
the Fermi-liquid regime in the BEC--fermion mixture.

\section{Acknowledgement}

D.H.S.'s work is supported by the NSF through a grant for the
Institute for Theoretical Atomic, Molecular and Optical Physics
(ITAMP) at Harvard University and Smithsonian Astrophysical
Observatory. E.T. would like to thank ITAMP for a visit in December,
2004. We are grateful to KITP, University of California at Santa
Barbara, for providing a stimulating environment during part of our
collaboration.

\end{document}